# Information and Data Quality in Spreadsheets

Patrick O'Beirne

Systems Modelling Ltd

Tara Hill, Gorey, Co. Wexford, Ireland

Tel +353-53-942-2294 Email pob@sysmod.com

**Abstract**

The quality of the data in spreadsheets is less discussed than the structural integrity of the formulas. Yet it is an area of great interest to the owners and users of the spreadsheet. This paper provides an overview of Information Quality (IQ) and Data Quality (DQ) with specific reference to how data is sourced, structured, and presented in spreadsheets.

## *Introduction*

Much of the focus on spreadsheet quality is naturally concerned with the formulas and their integrity. While most users and their managers are worried about the problems caused by Garbage In, Garbage Out (GIGO), their interest rarely goes further than a concern with accuracy. In fact there are many more data attributes that need to be covered in a serious review.

This paper has four main sections:

1) A brief outline of interest in Data and Information Quality

2) A review of the data attributes commonly described in the literature on data quality;

3) A review of papers and software tools;

4) Considerations specifically to do with spreadsheet data control

## *1. Information and Data Quality*

Like Eusprig, there is an International Association for Information and Data Quality - www.iaidq.org

The visitor statistics for the Eusprig 'Horror Stories' page www.eusprig.org/stories.htm show that people like to read about others' misfortune. There is a similar site on 'train wrecks' in Information Quality:

http://www.iqtrainwrecks.com/ An IQ Trainwreck is a problem that affects real people in the real world that has, at its heart, poor quality information or a failure to manage the quality of information. These can range from the inconvenience of dealing with poor customer service from poor quality data/information to the loss of life or limb that might arise if there is a failure to manage Information Quality appropriately.

## How expensive can bad quality data be?

In one case $125,000,000: the price of a Mars Climate Orbiter lost in space in September 1999. [9] The peer review preliminary findings indicate that one team used English units (e.g., inches, feet and pounds) while the other used metric units for a key spacecraft operation. Dr. Edward Weiler, NASA's Associate Administrator for Space Science said "The problem here was not the error, it was the failure of NASA's systems engineering, and the checks and balances in our processes to detect the error. That's why we lost the spacecraft."

English [8] points out that "Some think that by 'cleansing' or 'correcting' data they are improving information quality. Not true. Like manufacturing scrap and rework, data cleansing is merely rework to correct defects that would not be there if the processes worked properly. Data cleansing and correction are, simply put, part of the costs of nonquality data." He tells a story about an insurance company that had 80% of its claims recorded as for broken legs to distinguish the concept of 'valid' – that is, the users found a way to make data entry quickly pass the validation routine; from 'correct' – that is, it corresponds to reality.

## 2. Information Quality Attributes

Information quality can be measured using one or more of the following dimensions: [13]

| Dimension | The extent to which the information is or has … |
|---|---|
| Accessible | available, or quickly and easily retrievable |
| Accuracy | Accuracy can be thought of as freedom from mistake or error. Accuracy exists when reality and what is recorded as data are in agreement. |
| Appropriate Amount | the amount appropriate for the task at hand |
| Atomic | only one fact in a given field |
| Believable | regarded as true or credible, transparent (errors not hidden) |
| Complete | not missing and is of sufficient breadth and depth |
| Concise | compactly represented |
| Coverage | How much of what is available has been recorded. |
| Conformity | presented in the same format, eg dates |
| Consistent | values in the fields do not conflict (eg name=John gender=F) |
| Coherence | integrity, agreement with related data |
| Interpretable | in appropriate language, symbols, or units, and definitions clear |
| Meaning | the information recorded with a field agrees with the definition of the field |
| Objective | unbiased, unprejudiced, and impartial |
| Redundancy | Single instance is the ideal; sometimes redundancy is accepted for performance reasons |
| Relevant | applicable and helpful |
| Reputable | highly regarded in terms of its source or content |
| Secure | access is restricted appropriately to maintain its security, authentication, privacy, IPR, copyright, and legal or regulatory requirements |
| Timely | sufficiently current or up-to-date for the purpose. 'Float' is the lag between a fact being recorded in System A and it being passed to System B |
| Understandable | easily comprehended |
| Usability | ease of manipulation to apply to different tasks |
| Value | beneficial and provides advantages from its use; from the other point of view, the risk in not having important data |
| Validity | Passes edit controls which rule out impossible or unusable values or impose business policies |

**Table 1: Information Quality Attributes**

It will be noted that many of the above are not technical attributes, and result from process assurance around the collection, preservation, and presentation of the data. Table 2 shows data item attributes, and Table 3 shows some of the control and test processes available to the auditor.

Figure 1 is a fishbone (cause-and-effect) diagram from ' Data Quality: The Logistics Imperative' [10] illustrating the many contributory causes to poor data quality.

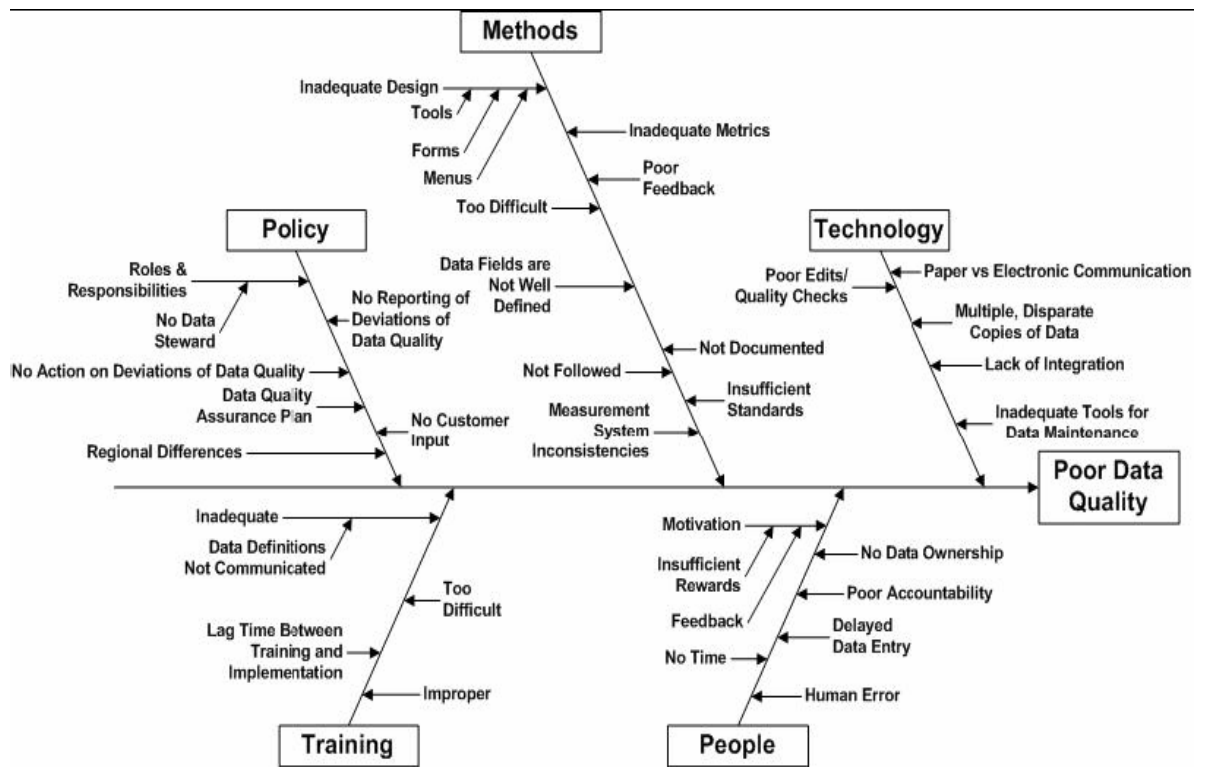

**Fig.1 Root Causes of Poor Data Quality**

| Data Item Attribute | Definition |
|---|---|
| Data type | Storage type used for the data element |
| Default value | If the user makes no entry the system enters this default value. |
| Error | Errors include transposition of digits and other keying mistakes. |
| Format | Presentation of the raw data in a form to aid comprehension |
| Missing value | What the system assumes if the value is empty |
| Null | Whether Null is allowed |
| Precision | The measure of the detail in which the quantity is expressed |
| Primary key | Must be unique |
| Range of values | Minimum to maximum valid values |
| Referential integrity | Primary and Foreign Keys must exist between parent and child records. |
| Restricted value list | List of possible valid values |
| Size | Length in characters or scale |
| Unit of measure | For quantities |

**Table 2: Data item attributes**

Attributes in table 2 are part of metadata, that is, data describing the context, content and structure of records and their management through time.

| Checks | Definition |
|---|---|
| Procedures | Are there defined procedures for processing the data and are they followed? |
| Controls | Authorisation and separation of duties |
| Audit trail | Auditing answers the question what was changed or viewed, by what user and on what date and time. |
| Audit checks | A re-check against business rules for example to reconcile two accounts; to detect whether there are many payments just below an authorisation threshold. |
| Archive | Secure access to backups |
| Tags | Data can be tagged or flagged to indicate non-conformity |
| Merge | Can sets of records be merged without contamination; or unmerged? Can different time series, scales, or types be integrated? |
| Missing records | How can you know when records are absent? |
| Process | Frequency and time characteristics |
| Sharing | What other systems have access, and to what level? |
| Source | Where did the data come from? |
| Use | Where is the data used? |
| Linked data | Automatic links between spreadsheets must be controlled for completeness, accuracy and appropriateness of data transfer. |
| Continuous | Are there gaps in some field sequences? |
| Duplicates | Are records duplicated? |
| Statistics | Measures such as Min/Bottom 10, Max/Top 10, Average, Frequency distribution; can indicate expected values and help in identifying outliers and unexpected or unlikely values. |
| Ambiguity | Is there more than one field of the same name with different meanings? |
| Error statistics | Statistics on the expected occurrence of random errors; for example transcription errors. |
| Calibration | Validating a measuring instrument against a standard |
| Sampling | Where there is more data than can be checked, a sample must be taken following standard statistical sampling techniques. |

**Table 3: Checks in Information and Data Quality processes**

## 3. Data Quality in the literature on spreadsheets

### Spreadsheet Modelling

Classically spreadsheets are considered modelling tools and much of the literature deals with questions of forecasting based on historical data or projections based on the expectations of marketing people. An example is Read and Batson [3] who refer to 'Data requirements'. Their focus is on modelling and therefore on the difficulties in producing appropriate data. To manage the data requirements they give an example of a data input list which specifies for each input item its Format, Units, Frequency of update, Status of its authority, Validation rule, Source, and other notes. The list

> • establishes the level of detail required in the input assumptions at an early stage, in a form that is easy to communicate;
>
> • highlights where special effort will be required in data collection – and assigns responsibility for producing the data; and
>
> • in the status column, lists the type of estimate that will be used for each input assumption. This can help identify the inputs that will be candidates for sensitivity analysis.

Read & Batson also state that if you are using a lot of the database functions, you probably should be using database software, because spreadsheets are weak for handling multi-dimensional data and large volumes. This advice is often repeated by IT professionals but rarely do users have the skills even in Microsoft Access to do this. The best one can expect to find is data structured as database tables in rows and columns with unique headings.

### The effect of data errors on forecasting models

Ballou [4] emphasize the sensitivity to errors in the denominator of a ratio. "One consequence of erroneous data is the multiplier effects which can arise when incorrect data undergo many and sequential manipulations. … Our analysis focuses on the relative magnitude of errors throughout the spreadsheet where relative error is defined by the coefficient of variation. … As would be anticipated, aggregation significantly dampens input error."

"Although it is widely known that the expected value of a ratio is not equal to the ratio of expected values, such ratios are often a major output of spreadsheet analysis." This is referred to by Savage [5] as the Flaw of Averages: "Plans based on average conditions are wrong on average". To avoid this, more sophisticated forecasters use products such as @RISK and Analycorp's XLSim.

On the other hand, errors may actually better suit forecasts. For construction of the linear regression models, the historical data is referred to as the training set. Klein & Rossin [14] conclude "For errors in training data, it is demonstrated that the predictive accuracy of a linear regression model built to forecast the [Net Asset Value] of mutual funds is better when errors exist in training data than when training data are free of errors. All of the scenarios with errors have predictive accuracy significantly better than the base case scenario without data errors. This is a significant contribution to the literature on data quality because this is the first research finding on the effect of errors in training data on linear regression models. This finding demonstrates that perfectly accurate data may not always provide the best forecast."

## Data Manipulation

More recently, it is recognised that much present use of spreadsheets is as data manipulation and reporting [16] tools used to bypass the controls around IT development. This ad hoc integration, transformation, or simple cobbling together is done by the user to get what they need when they need it. This gives rise to many extracted copies of corporate data as imports or query links in spreadsheet files. These personal data stores are often referred to as 'data shadows' or 'silos' or 'spreadmarts' giving rise to 'multiple versions of the truth'. The data massaging that they perform are an example of stovepipe systems. (Wikipedia []: In engineering and computing, a stovepipe system is a legacy system that is an assemblage of inter-related elements that are so tightly bound together that the individual elements cannot be differentiated, upgraded or refactored. The stovepipe system must be maintained until it can be entirely replaced by a new system.)

With 1 million rows and 16,000 columns in Excel 12, we can expect much more data analysis to be carried out in spreadsheets in the future.

## Computer Aided Audit Tools & Techniques (CAATTs)

SpACE, the spreadsheet auditing tool referred to by Butler[1], can check lists of data for duplicates and attempt to identify the numbers that make up a particular total. ACL (Audit Command Language) and IDEA (Interactive Data Extraction and Analysis) provide functions plus libraries of prewritten queries that help sample and examine data sets. Utilities such as ActiveData for Excel, the Cirrus Data Consistency Checker, and WizRule Data Mining explore large data sets and spreadsheets. The following is a list of relevant websites:

> http://en.wikipedia.org/wiki/Data_analysis_(information_technology))
>
> http://www.auditware.co.uk  Auditware intend to bring SpACE to market
>
> http://www.informationactive.com/ad   ActiveData for Excel
>
> http://www.informationactive.com/data/attachments/fraudsoftware.PDF Comparing Best Software for Fraud Examinations by Rich Lanza
>
> http://www.cirrussoftware.com/datacleansingtool  Data Cleansing Tool (in development)
>
> http://www.acl.com Audit Command Language
>
> http://www.caseware-idea.com  IDEA - Interactive Data Exploration and Analysis
>
> http://www.picalo.org  Open Source Python-Based Data Analysis Platform
>
> http://www.wizsoft.com  WizSoft Data Mining

## Typical functions of CAATTS

Match and Merge: Combines columns from sheets where rows are matched by some comparison operator.

Compare: compares two sheets. Is the copy of the data in the spreadsheet consistent with the original data stored in the source database?

Extract: Separate a sheet into multiple sheets based on values in a column. Sampling.

Generate: Fill cells with random, fixed or incremental values, characters, dates, or numbers

Convert: transform or reformat data formats or data types. Did the spreadsheet correctly process the format and data types of the data at the time?

Group: Subtotals, Top/Bottom Items, Date Aging, Stratification by bands, Cross-tabulation

Statistics: Descriptive Statistics, Summary

Duplicates: duplicated rows (are primary keys still unique?)

Gaps: missing rows, data items missing (empty cells), or invalid

Find: suspicious data (all the 9s, 01/01/01, and similar)

Spell-check: are there any spelling mistakes?

Benford's analysis: used to detect fraud from the pattern of digits where amounts have been invented.

## Visualisation

Graphing data has always been a common way to detect outliers. Now data mining tools offer a more powerful way to find patterns and breaks in patterns.

Grossman [11] comments "We are not aware of any research efforts to apply information visualization techniques specifically toward data quality." In his Java-based data visualization tool DaVis, he used the Color Brewer system, a tool developed by Cindy Brewer (1994) to pick a set of colours that indicate sequential differences between records. The aim is to find invalid and missing values, discovering zeros and other suspicious values such as 99 or 99999, identifying duplicated rows and columns, and detecting the differences between two sets of data.

## Considerations specific to spreadsheets

Read and Batson [3] recommend that inputs be clearly distinguished from formulas to reduce the risk that parts of the input data are overlooked. Inputs can be separated either physically on the sheet, or by clear labelling and colouring of the spreadsheet.

Self-checking formulas can be used:

> • on balancing financial statements, such as a balance sheet or a sources and applications of funds statement;
>
> • when equivalent data is provided from more than one source, for example when a total revenue is both input and calculated from individual revenue lines; and
>
> • when it is possible to perform a calculation in two equivalent but different ways.

Butler[1] describes the process used by a tax inspector concerned not just with the integrity of spreadsheet formulas but with the correctness of the data that is represented as a company's return of value-added tax (US: sales tax). He says "Base numbers are checked to ensure they match source documents or other data." And [2] "Good documentation should make clear statements of what standing data constants (e.g. tax, duty, interest and exchange rates) are used and where they are held."

In the context of data control, Butler states: "In common with all computer applications, the accuracy of the results of processing depend on the completeness, accuracy, timeliness and authorisation / appropriateness of the data. Even when a properly validated specification has been verified as implemented correctly, with all domain and arithmetic issues correctly handled, the GIGO (for younger readers, Garbage in, Garbage Out) principle still applies. The auditor must therefore ask what controls are built into the application to ensure that:

·       all relevant data are input,
·       no irrelevant or inappropriate data are input,
·       data are input accurately,
·       data are input for process at the correct time, and
·       if data is passed from one file or worksheet to another, adequate controls are in place to ensure the completeness and accuracy of the transfer"

Interestingly, Howe and Simkin [6] found mean error-detection rates of as high as 72% for data-entry errors compared to 54% for formula errors.

Kruck[7] refers to data quality, data validation, and test data. Test data is rarely created with sufficient variety to cover all the possible logical conditions in a spreadsheet. The interested reader is referred to similar discussions on code coverage in software testing.

Winston [15] in his book on business modelling provides a large number of examples that exercise the Data Analysis functions of Excel 2007.

The Spreadsheet Safe syllabus [12] covers many of the items referred to here, such as "Set out conventions used", "Isolate constants", "Import CSV & validate", "Validate links", "Check for missing input values", "Use IF to test values expected", "Review for data type mis-entry", "Apply conditional formatting to highlight errors", "Apply validation criteria".

O'Beirne [17] on Spreadsheet Check and Control and Jelen & Dowell [18] on Excel for Auditors both cover techniques for data quality control and auditing.

There are several ways to present the results of tests on a spreadsheet:

- A list of items. For example, a list of defined names or data validation rules. This is most commonly used when you want to document the work or report issues to others.

- A selection of cells. You can move around the selection using the Tab or Enter keys. This is most commonly used when you want to edit the found cells. Excel is restricted to 8192 discrete (non-contiguous) areas in a selection.

- Highlighted cells. For example, exceptions on a sheet might be coloured red by a conditional format or a VBA macro. This is most commonly used when you want to explore the worksheet looking for patterns or breaks in patterns. It is less successful when there are a few highlighted cells in a large range as it depends on you seeing the highlight colour.

These methods can be combined. For example, you can first select cells of interest then apply a colour to them, or use a macro to output the list of addresses and attributes (such as values or formulas) to a new sheet.

Appendix 2 presents a list of such techniques for applying data quality tests in Excel.

Auditors will use continue to use software tools to sample and interrogate large data sets, and for the kinds of pattern detection that are not built into Excel. However, Excel 2007 is catching up and many users will be satisfied with the facilities built in to it and add-ins such as the SQL Server Data Mining Add-Ins for Office 2007.

In conclusion, it is possible to apply many data quality controls and tests into spreadsheets. To assure information quality will remain a matter for process assurance and management.

## *References*

[1] R. Butler. 2000a. "Is This Spreadsheet a Tax Evader?" Proceedings of the 33rd Hawaii International Conference on System Sciences, pp. 1-6.

[2] R. Butler. 2000b. "Risk Assessment for Spreadsheet Developments: Choosing Which Models to Audit." H. M. Customs and Excise, UK.

[3] Spreadsheet Modelling Best Practice by Nick Read and Jonathan Batson of Business Dynamics April 1999, http://www.eusprig.org/smbp.pdf  accessed Apr 24, 2008.

[4] Ballou, D.P., Pazer, H.L., Belardo, S., and Klein, B. (1987) "Implications of Data Quality for Spreadsheet Analysis," Data Base, V. 18, N. 3, Spring, pp. 13-19.

[5] The Flaw of Averages and What to Do About It, Dr. Sam Savage, to be published 2008. http://flawofaverages.com/ accessed Apr 24, 2008.

[6] H. Howe and M. Simkin. 2006. "Factors Affecting the Ability to Detect Spreadsheet Errors." Decision Sciences Journal of Innovative Education. Vol. 4:1, pp. 101-122.

[7] S. Kruck. 2006. "Testing Spreadsheet Accuracy Theory." Information and Software Technology, Vol. 48, pp. 204-213.

[8] Seven Deadly Misconceptions about Data Quality, Larry P. English, http://www.fhwa.dot.gov/policy/ohpi/dataquality.htm  accessed Apr 24, 2008.

[9] Mars Climate Orbiter Failure Board Releases Report, http://mars.jpl.nasa.gov/msp98/news/mco991110.html accessed Apr 24, 2008.

[10] Data Quality: The Logistics Imperative, Elaine S. Chapman http://www.eccma.org/.../2006Conf/Data%20Quality%20and%20Spend%20Analysis/Data_Quality-Spend_Analysis-Chapman.ppt

[11] DaVis: A tool for Visualizing Data Quality, Grossman 2005 http://www.rgrossman.com/dl/proc-095.pdf accessed Apr 24, 2008.

[12] Spreadsheet Safe training and certification of user competence http://www.spreadsheetsafe.com  accessed Apr 25, 2008.

[13] Data Quality Assessment. Leo L. Pipino, Yang W. Lee, and Richard Y. Wang Communications of the ACM April 2002/Vol. 45, No. 4ve. 211. http://web.mit.edu/tdqm/www/tdqmpub/PipinoLeeWangCACMApr02.pdf   Apr 24, 2008.

[14] Data Quality in Linear Regression Models: Effect of Errors in Test Data and Errors in Training Data on Predictive Accuracy, Barbara D. Klein & Donald F. Rossin, http://inform.nu/Articles/Vol2/v2n2p33-43.pdf  accessed Apr 25, 2008.

[15] Excel 2007 Data Analysis and Business Modelling, Wayne L Winston, Microsoft Press 2007, ISBN 0-7356-2396-1

[16] Excel Advanced Report Development, Timothy Zapawa, Wiley Publishing 2005, ISBN 0-7645-8811-7

[17] Spreadsheet Check and Control, Patrick O'Beirne, Systems Publishing 2005, ISBN 1-905404-00-X

[18] Excel for Auditors, Bill Jelen and Dwayne K Dowell, Holy Macro! Books ISBN 1-932802-16-9

## *Appendix 1*

## Audit Reports on Data Quality involving spreadsheets

http://publications.environment-agency.gov.uk/pdf/GEHO0207BMWI-e-e.pdf  Environment Agency Data Quality Report for LATS City of London 2007

A spreadsheet […] is formatted such that the new organics recycling scheme does not contribute to the sum of waste recycled. There were also transcription errors in taking data onto the same spreadsheet from the weighbridge attendant's reports and re-entering it into the summary spreadsheets.

http://www.knowsley.gov.uk/shared/javascript/common/logging_script.php?transtype=Providing+Information&pidno=0&path=/resources/203531/audit0506.pdf

Knowsley Metropolitan Borough Council 2005-06, Report to those charged with governance (PWC)

Errors in the use of spreadsheets have resulted in two significant errors being identified which highlight a significant control weakness at the point of review.

The draft financial statements presented for audit contained […] elements overstated by £9,159,000 as a result of errors in the use of spreadsheets.

http://www.midsussex.gov.uk/Nimoi/sites/msdcpublic/resources/item%207,%20appendix%201.pdf

Data Quality Review Mid Sussex District Council Audit 2007-08

We tested a sample of twenty survey records from the spreadsheet, used to calculate the PI, back to the supporting forms and noted that 22 out of 800 entries were incorrect. […] For these reasons we have concluded that the PI is not fairly stated.

http://committee.west-lindsey.gov.uk/comm_mins/documents/ASC/Reports/appendix%20ASC0021R.pdf

Data Quality West Lindsey District Council Audit 2007-08

The Council undertook additional testing which confirmed errors in the transfer of data from the field officer forms through to the DEFRA spreadsheet equating to an error rate of 14 per cent over the year.

http://www.oecd.org/dataoecd/26/12/18989820.ppt

Introducing a Quality Management System (or finding the positive side to an error: the UK CPI experience)

Published headline inflation rate for March and May 1995 understated by 0.1% (inflation around 3.5%) UK RPI (CPI) never revised = publicly stated policy. The error occurred in a spreadsheet used to process a centrally collected price.

http://www.eiminstitute.org/current-magazine/volume-1-issue-5-july-2007-edition/a-foundation-for-data-quality

"While transforming the data from the spreadsheet into the corporate accounting system, a decimal point had been misplaced resulting in a difference in the millions of dollars. In the operational systems the monetary values were recorded in actual currency units and the spreadsheet recorded the data in thousands of dollars. The decision was made to record the data in the global data warehouse as collected from the operational systems and to not restate the financial statements because the difference was not material."

## *Appendix 2*

This appendix describes some techniques used by auditors to analyse data. It does not detail all the data analysis functions of Excel which one can discover by consulting the help files or the many books [15-18] on the topic; eg Spell-check, Find, Frequency function, Cell function, Data Validation, Charting, Sorting, Grouping, Subtotalling, Pivot Tables, etc.

## Excel features

Data Validation is the most obvious form of data control. It is often pointed out that it can be too easily overwritten by copy and paste. Much ingenuity has been expended on ways to get users to paste values only, or disable the several ways in which Excel can paste data (Ctrl+V, right click menu, edit paste menu items, drag & drop).

Conditional Formatting is a built-in feature of Excel that allows formatting such as colours, borders, patterns, and fonts to be applied to cells depending on a number of criteria. In Excel 2007, conditional formatting can contain up to sixty-four conditions, but in earlier versions of Excel, only three conditions are supported.

Pivot Tables and their associated Pivot Charts are a popular way to carry out cross-tabulation in Excel. Before Excel 12 (Excel 2007), spreadsheets were limited to 65535 rows. However, the data queries on which PivotTables are based have no such limit, so it is possible to query unlimited data sets as long as the result set does not exceed 65535 rows.

Excel 2007 has multiple selection in AutoFilters, can sort or filter by colour, and provides "quick filters" for specific data types. It supports Microsoft SQL Server 2005 Analysis Services and the new cube functions allow you to build a custom report from an OLAP database.

## Excel tips and techniques

1) To select input cells with missing values, first select the formula(s) whose inputs you want to check. Then on the Edit menu click Goto, click the Special button, select Precedents, select All levels. The shortcut is Ctrl+{. Without changing the selection, On the Edit menu click Goto, click the Special button, select Blanks.

2) To select all input cells, do as above but in the final step select Constants rather than Blanks. Now you can format the selection, or give it a name, or unlock the input cells prior to protecting the sheet.

3) When you have an input area selected, you can enter the same data (eg zero) into all of the cells by typing it and pressing Ctrl+Enter.

4) To highlight invalid entries that fail data validation rules, point to Formula Auditing on the Tools menu, and then click Show Formula Auditing Toolbar. Click Circle Invalid Data, which draws red circles around the invalid cells. There is no menu method to select the cells. A VBA macro to do so is provided in the SelectInvalid sub in the sample file.

5) To extract unique data to a separate sheet, select the data first. If you choose more than one column the entire row is used to determine uniqueness. On the Data menu click Filter, click Advanced Filter, select Copy to another location, enter the location in the Copy to box, check Unique records only, click OK.

6) To count the duplicate values in Column A, enter =COUNTIF(A:A,A1) in cell B1 and copy down.

7) To highlight duplicates with conditional formatting, first select the cells to be checked. On the Format menu select Conditional Formatting. For Condition 1, select Formula Is. In the Formula box enter =COUNTIF(A:A,A2)>1 where A2 is for example the first cell of the selection. Click the Format button and select for example on the Pattern tab a Cell shading colour of red. Click OK, and OK again.

8) To implement validation to prevent duplicates in data entry, select the range of cells that you wish to restrict, and chose Validation from the Data menu. Choose Custom from the Allow list, and enter the following formula: =COUNTIF(A:A,A1)=1 where A1 is for example the first cell in the selection.

9) To find duplicate items between two columns A and B, in column C enter and copy down the formula =IF(ISERROR(MATCH(B1,A:A,0)),"",B1)

10) To find gaps, sort the data and add a column that calculates the difference between successive items. Set an Autofilter on this column and now differences of 0 show duplicates, 1 show sequential items, 2 or more show gaps.

## Comparing sheets

If the structure of both sheets is different, you may first have to compare them assuming they are the same, and when you find the differences, insert rows or columns to make them line up again from that point. Another approach is to export each sheet as a space delimited text and then use a standard DIFF utility on the text files.

If the structure of both sheets is the same and only the values are different, then you only need to compare values as follows - first backup the original file!

Select all cells in one sheet and copy. On the other sheet select Edit > Paste Special and select Subtract. This subtracts the numeric constants and does not affect non-numerics.

To remove the zeroes and leave just the nonzeroes, on the Edit menu click Replace, enter 0 in the Find What box, clear the Replace With box, use the Options button to show the options and check Match Entire Cell Contents, and then click the Replace All button.

To locate the remaining cells, on the Edit menu click Goto, click the Special button, select Constants, select Numbers only, click OK.

Another method is to insert a new sheet and put a formula in every cell comparing the same cell in the other two sheets. Then simply find all FALSE values.

There are of course many commercial tools for this, such as ScanXLS, John Walkenbach's Power Utility Pack, Spreadsheet Professional, Synkronizer, etc

Web site references follow.

## Web references

http://www.sysmod.com/scanxls.htm ScanXLS can inventory spreadsheets, report on properties and errors, and compare two workbooks.

http://www.dailydoseofexcel.com/archives/2006/06/24/handle-poor-external-data-quality/ Measure & Evaluate The Quality of The Data, accessed Apr 24, 2008. Wallentin and others discuss data quality in an Excel user group.

http://www.vbaexpress.com/kb/getarticle.php?kb_id=966

Compare two ranges and output a list of all the differences and their locations.

http://www.erlandsendata.no/english/index.php?d=envbawscomparews

Compare two worksheets

http://www.formulasoft.com/vba-code-compare.html

Free addin to compare VBA code. Site has commercial workbook comparison utility.

http://www.cpearson.com/excel/Duplicates.aspx

Techniques for dealing with duplicate items in a list of data.

http://office.microsoft.com/en-us/excel/HA011039151033.aspx

Use Excel to compare two lists of data

http://office.microsoft.com/en-us/excel/HA010346261033.aspx

Delete duplicate rows from a list in Excel. A duplicate row (also called a record) in a list is one where all values in the row are an exact match of all the values in another row.

http://www.cpearson.com/excel/deleting.htm

Macros (DeleteBlankRows, DeleteRowOnCell, and DeleteDuplicateRows) which will delete all blank rows or all duplicate rows from a range of rows in a worksheet.

## *End of appendices*